\documentclass{IEEEtran}

\ifCLASSINFOpdf
   \usepackage[pdftex]{graphicx}
\else
   \usepackage[dvips]{graphicx}
\fi
\usepackage{amsfonts}
\usepackage[cmex10]{amsmath}
\usepackage{wrapfig}
\usepackage{caption}
\usepackage{xcolor}
\usepackage{todonotes}
\usepackage{float}

\ifCLASSOPTIONcompsoc
 \usepackage[caption=false,font=normalsize,labelfont=sf,textfont=sf]{subfig}
\else
 \usepackage[caption=false,font=footnotesize]{subfig}
\fi
\usepackage{float}
\ifCLASSOPTIONcompsoc
 \usepackage[caption=false,font=normalsize,labelfont=sf,textfont=sf]{subfig}
\else
 \usepackage[caption=false,font=footnotesize]{subfig}
\fi

\floatstyle{ruled}
\newfloat{model}{thp}{lop}
\floatname{model}{Model}

\usepackage{pifont}
\newcommand{\cmark}{\ding{51}}%
\newcommand{\xmark}{\ding{55}}%

\makeatletter
\let\old@ps@headings\ps@headings
\let\old@ps@IEEEtitlepagestyle\ps@IEEEtitlepagestyle

\makeatother

\usepackage{booktabs}
\usepackage{lastpage}
\usepackage{algorithmic}
\usepackage[ruled,linesnumbered,noresetcount,vlined]{algorithm2e}
\usepackage[noadjust]{cite}
\usepackage{multirow}
\usepackage{makecell}

\DeclareMathOperator{\Var}{Var}  
\DeclareMathOperator{\SAVar}{SAVar}  
\DeclareMathOperator{\Intermittency}{Imcy}  

\DeclareMathOperator{\tr}{tr}  

\begin{document}

%
\title{Asset Bundling for Wind Power Forecasting}

\author{
\IEEEauthorblockN{Hanyu Zhang, Mathieu Tanneau, Chaofan Huang, V. Roshan Joseph, Shangkun Wang, Pascal Van Hentenryck}
\IEEEauthorblockA{
Georgia Institute of Technology, Atlanta, USA\\
\{hzhang747, chaofan.huang, roshan, sk$\_$wang\}@gatech.edu, \{mathieu.tanneau, pascal.vanhentenryck\}@isye.gatech.edu
}
}


\maketitle


\begin{abstract}
The growing penetration of intermittent, renewable generation in US
power grids, especially wind and solar generation, results in increased
operational uncertainty.
In that context, accurate forecasts are
critical, especially for wind generation, which exhibits large
variability and is historically harder to predict.
To overcome this challenge, this work proposes a novel Bundle-Predict-Reconcile (BPR) framework that integrates asset bundling, machine learning, and forecast reconciliation techniques.
The BPR framework first learns an intermediate hierarchy level (the bundles), then predicts wind power at the asset, bundle, and fleet level, and finally reconciles all forecasts to ensure consistency.
This approach effectively introduces an auxiliary learning task (predicting the bundle-level time series) to help the main learning tasks.
The paper also introduces new asset-bundling criteria that capture the spatio-temporal dynamics of wind power time series.
Extensive numerical experiments are conducted on an industry-size dataset of 283 wind farms in the MISO footprint.
The experiments consider short-term and day-ahead forecasts, and evaluates a large variety of forecasting models that include weather predictions as covariates.
The results demonstrate the benefits of BPR, which consistently and significantly improves forecast accuracy over baselines, especially at the fleet level.
\end{abstract}


\thanksto{\noindent This research was partly supported by NSF award 2112533 and ARPA-E PERFORM award AR0001136.}

\section{Introduction}
\label{sec:introduction}

The sustained growth of wind power generation in US power grids is
causing operational and reliability challenges for Transmission System
Operators (TSOs). In particular, the intrinsic variability and
intermittency of wind resources creates significant operational
uncertainty, which must be managed to ensure safe and reliable
operations.  This, in turn, requires accurate wind power forecasts at
various temporal and geographical scales.  Indeed, market participants
use wind power forecasts at the wind farm level to inform their
participation in day-ahead and real-time markets, while TSOs primarily
use aggregated, fleet-wide forecasts to operate the grid.  Forecasts
must be accurate at both levels while remaining consistent, i.e.,
the sum of individual forecasts should be equal to the forecast for
the overall fleet. As a result, this paper considers the task of
producing high-dimensional, short-term and day-ahead, wind power
forecasts that satisfy three requirements:
\begin{enumerate}
    \item the forecasts predict the output of individual wind farms and the total output of the fleet;
    \item the forecasts are {\em consistent}, i.e., the total forecast is the sum of the individual forecasts;
    \item the forecasts are accurate both at the individual level and at the fleet level.
\end{enumerate}
To carry this task, the paper proposes a novel hierarchical approach,
{\em the Bundle-Predict-Reconcile} (BPR) framework, which (1) learns
an additional forecast level ({\em the bundles}) between the assets
and the fleet, (2) predicts outputs at the asset, bundle, and fleet
levels, and (3) {\em reconciles} all levels for ensuring consistency. The
rest of this section surveys the relevant literature and presents the
paper's contributions.

\subsection{Related Work}
\label{sec:intro:literature}

Because of its importance for power systems operations, wind power
forecasting has received substantial attention. Readers are referred
to \cite{Soman2010_WindPowerForecastingReview,Tao2020_EnergyForecastingReview}
for comprehensive reviews. For ease of presentation, the paper focuses
on deterministic forecasts but BPR naturally extends to probabilistic
settings. See \cite{Zhang2014_ProbabilisticWindForecastReview} for an
overview of probabilistic wind forecasting.

Wind power forecasting methods broadly fall under two categories:
physical and statistical models.  On the one hand, physical
methods \cite{Xu2015_ShortTermWindPowerForecastingWeather,Yan2019_AdvancedWindPowerPrediction}
combine weather-based wind speed predictions with wind power
curves \cite{WPcurve} to produce wind power forecasts.  On the other
hand, statistical methods use historical data to train data-driven
models.  These include traditional time series forecasting models,
such as Auto-Regressive Moving Average (ARMA) and its
variants \cite{Soman2010_WindPowerForecastingReview}.  More recently,
Machine Learning (ML) models have been applied to wind power
forecasting, e.g., Artificial Neural
Networks \cite{Soman2010_WindPowerForecastingReview}, recurrent neural
networks \cite{NIU2020117081,Salinas2020_DeepAR}, and
transformer-based
architectures \cite{TCN_WindPower,TCNPaper,lim2019temporal}.  A
comparison of several deep learning models for wind power forecasting
is presented in \cite{MASHLAKOV2021116405}.  Statistical models are
typically better at short-term ($\leq$6 hours) predictions, but are
outperformed by physical models for longer
horizons \cite{Soman2010_WindPowerForecastingReview}.  Current
state-of-the-art approaches thus combine physics-based weather
predictions with data-driven
models \cite{Sweeney2020_ForecastingRenewableEnergy,Tao2020_EnergyForecastingReview}.

Wind power forecasts are often produced in a hierarchical fashion,
with different models predicting different levels of the hierarchy.
For instance, one model predicts the output of individual wind farms,
and another model predicts the fleet's total output.  Forecast
reconciliation \cite{Hyndman2011_ForecastHierarchicalTimeSeries,Shanika2019_MinT}
can then be used to ensure consistency of hierarchical wind
forecasts \cite{Zhang2018_OptimalReconciliationWindPower,Bai2019_DistributedReconciliation,DiModica2021_OnlineWindForecastReconciliation,Stratigakos2022_E2EHierarchicalForecasting,Hansen2023_ReconciliationWindPowerForecasts}.
Zhang et al. \cite{Zhang2018_OptimalReconciliationWindPower} compare
the performance of several reconciliation approaches for very
short-term ($\leq$1 hour) predictions.  Bai et
al \cite{Bai2019_DistributedReconciliation} propose a distributed
algorithm to solve the reconciliation problem, and an online
reconciliation approach is presented
in \cite{DiModica2021_OnlineWindForecastReconciliation}.  An
end-to-end learning approach for hierarchical wind power forecast is
introduced in \cite{Stratigakos2022_E2EHierarchicalForecasting} with a
special focus on handling missing values.  Hansen et
al. \cite{Hansen2023_ReconciliationWindPowerForecasts} find that
hierarchical reconciliation leads to improved day-ahead wind
forecasts, especially for total-level predictions.  In all these
works \cite{Zhang2018_OptimalReconciliationWindPower,Bai2019_DistributedReconciliation,DiModica2021_OnlineWindForecastReconciliation,Stratigakos2022_E2EHierarchicalForecasting,Hansen2023_ReconciliationWindPowerForecasts},
the hierarchical structure is fixed and given a priori.  In contrast,
the BPR framework proposed in this paper improves the forecast
accuracy by \emph{learning a hierarchical structure through asset
bundling.}
    
Asset bundling is a popular technique in portfolio optimization,
tracing back to the mean variance model of
Markowitz \cite{markowitz1952ps}.  In the context of power systems, a
similar approach is known as the geographical smoothing effect,
wherein the aggregated output of multiple wind farms exhibits lower
variability \cite{katzenstein2010bundling,bandi2017bundling,shahriari2017bundling,VariationAnalysis}.
The geographical smoothing effect has been employed in prior
literature on optimal design, planning, and re-powering of wind
farms \cite{hansen2004bundling,drake2007bundling,roques2010bundling,degeilh2011bundling,thomaidis2016bundling,santos2017bundling}.
References \cite{grothe2011bundling,ma2013bundling,gersema2018bundling,vinel2019bundling,han2022bundling}
use a similar approach, but consider a conditional value at risk
(CVaR) objective.  {\em Unlike the above portfolio selection-based
approaches, the paper proposes to use asset bundling to improve the
quality of wind power forecasts.}

\subsection{Contributions}
\label{sec:intro:contribution}

The BPR framework is a three step-approach that first learns a
prediction hierarchy by bunding assets, before forecasting time series
at the asset, bundle, and fleet levels, before reconciling the
forecasts at all levels. A key innovation of the BPR framework is its
ability {\em to learn the ``best'' bundles}. In other words, BPR is
not given a forecasting hierarchy; rather it learns a hierarchy by
choosing the bundles that will help improve the asset and fleet
forecasts. As a result, BPR can be viewed as a multi-task learning
approach: it introduces an auxiliary learning task, i.e., forecasting
the bundles, to help the main learning tasks. A second innovation of
BPR is the criteria used to select the bundles, leveraging the
spatio-temporal correlations that exist in time series of wind power.

The paper presents a comprehensive evaluation of BPR that considers
both short-term and day-ahead predictions, includes weather forecasts
as covariates, and evaluates a large variety of forecasting models for
the individual prediction tasks. Experimental results for
industry-size test cases from NREL \cite{nrelperformdata} clearly show the benefits of
BPR. They also give significant insights on the impact of bundling and
ML architecture on the quality of the forecasts. The
learning architectures include recurrent neural networks, transformer
models, temporal convolution transformers, and temporal fusion
transformers.  The bundling criteria include geographical distance,
seasonal-adjusted variance, and the intermittency index.

In summary, the paper makes three main contributions:
\begin{enumerate}
\item It introduces a new approach, Bundle-Predict-Reconcile (BPR), that learn
a forecasting hierarchy to improve the prediction accuracy at the asset and
fleet level.

\item It proposes new bundling criteria that exploit the spatio-temporal structure
of the time series for wind power.

\item It evaluates BPR on several forecasting tasks (short-term and day-ahead),
numerous learning architectures, and covariate configurations,
demonstrating its state-of-the-art performance for both short-term and
day-ahead wind power forecasts. In particular, BPR improves
fleet-level accuracy by about 25\% over a strong baseline in
short-term forecasting and about 10\% over the best models in
day-ahead forecasting.
\end{enumerate}

The rest of the paper is organized as follows.
Section \ref{sec:formulation} presents the problem formulation and
summarizes the proposed approach.  Section \ref{sec:bundling} presents
the proposed asset-bundling methodology.
Section \ref{sec:forecasting} describes the hierarchical forecasting
and reconciliation framework, and the ML architectures used in the
paper.  Section \ref{sec:experiments} presents numerical experiments,
and Section \ref{sec:conclusion} concludes the paper.

\section{Problem Formulation}
\label{sec:formulation}

\newcommand{\N}{\mathcal{N}}
\newcommand{\J}{\mathcal{J}}
\newcommand{\T}{\mathcal{T}}
\renewcommand{\H}{\mathcal{H}}
\newcommand{\W}{\mathcal{W}}

\newcommand{\x}{\mathbf{x}}
\newcommand{\xdot}{\mathbf{\dot{x}}}
\newcommand{\xhat}{\mathbf{\hat{x}}}
\newcommand{\z}{\mathbf{z}}
\newcommand{\zdot}{\mathbf{\dot{z}}}
\newcommand{\zhat}{\mathbf{\hat{z}}}

Let $\mathcal{N} \, {=} \, \{1, ..., N\}$ denote the set of wind farms
in the system.  The output of wind farm $i \, {\in} \, \{1, ..., N\}$
at time $t \, {\in} \, \mathbb{Z}$ is denoted by $\x_{i, t}$.
The following notations are used in the paper for $i \in \mathcal{N}$,
$t \in \mathbb{Z}$, and $\T \subseteq \mathbb{Z}$:
\begin{align*}
    \mathbf{x}_{t} 
        &= \left( \x_{i, t} \right)_{i \in \mathcal{N}} \in \mathbb{R}^{N}\\
    \mathbf{x}_{\T} 
        &= \left( \mathbf{x}_{\tau} \right)_{\tau \in \T} \in \mathbb{R}^{N \times |\T|}\\
    \mathbf{x}_{i, \T} 
        &= \left( \x_{i, \tau} \right)_{\tau \in \T} \in \mathbb{R}^{|\T|}
\end{align*}

\subsection{Wind Power Forecasting}
\label{sec:formulation:wind_forecast}

Let $t \, {\in} \, \mathbb{Z}$ denote the current time period, and
denote by $\H \, {=} \, \{t{-}H{+}1, ..., t\}$ and $\T \, {=} \,
\{t{+}1, ..., t{+}T\}$ the \emph{historical} and \emph{forecast}
windows, respectively.  Given past observations $\mathbf{x}_{\H}$, the
goal of wind power forecasting is to produce a forecast
$\xhat_{\mathcal{T}}$ that is as close as possible to the actual
(future) realization $\mathbf{x}_{\T}$.  While the paper considers
point forecasts for ease of presentation, the BPR methodology
naturally extend to probabilistic forecasting.

TSOs and market participants use wind power forecasts as input for
market-clearing and to conduct reliability studies.  In that context,
it is important to note that spatial correlations (between wind farms)
and temporal correlations (between time periods) have an impact on
congestion management and flexibility (ramping) requirements.
Therefore, forecasts should capture spatio-temporal dynamics, which is
achieved by jointly predicting all $N\times T$ components of
$\xhat_{\mathcal{T}}$.  The paper considers joint forecasts for
hundreds wind farms and time horizons containing 24 to 48 time steps,
giving prediction tasks with over 10,000 output dimensions. Such large
output dimensions create scalability challenges and have an adverse
impact on forecast accuracy.

\subsection{Overview of BPR}
\label{sec:formulation:asset_bundling}

\begin{figure}[!t]
    \centering
    \subfloat[%
        Vanilla wind power forecasting. Left: the initial two-level (wind farm, total) hierarchy. Middle: a model is trained to predict the wind farm and total level. Right: the resulting forecasts may not be consistent.
    ]{%
        \includegraphics[width=\columnwidth]{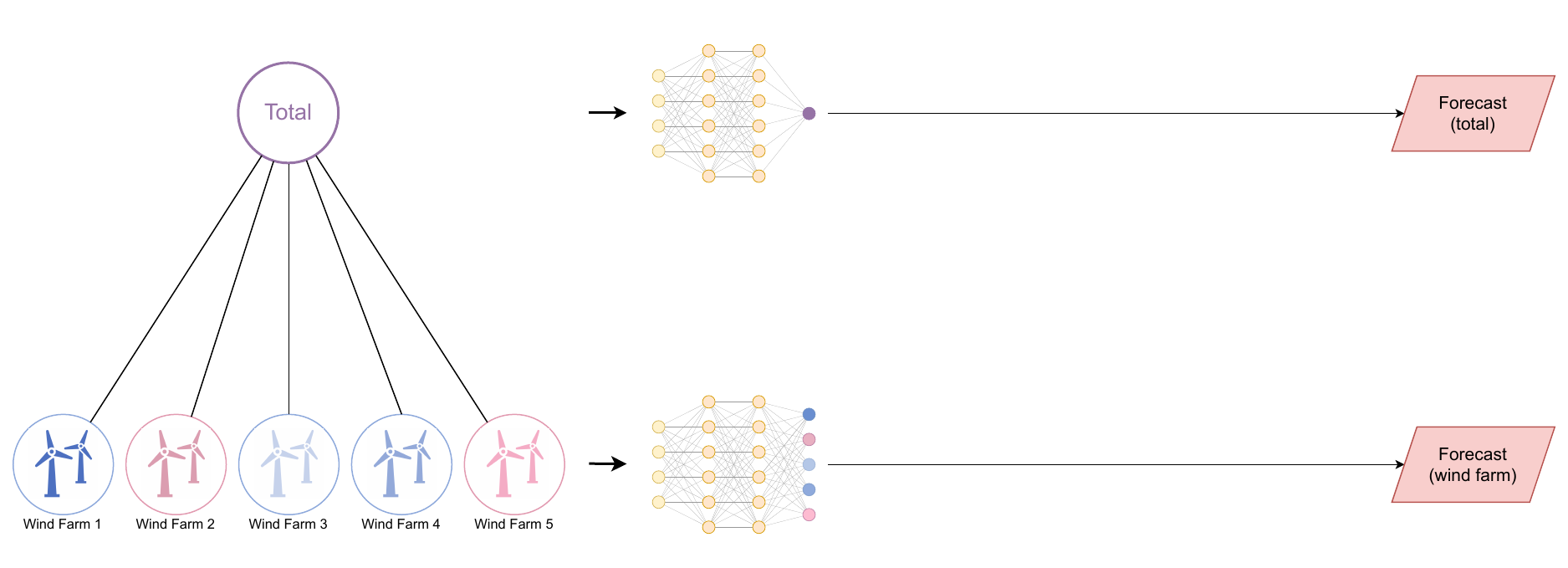}
        \label{fig:overview:vanilla}
    }\\
    \subfloat[%
        The proposed Bundle-Predict-Reconcile (BPR) framework. Left: wind farms are grouped into bundles. Middle: a model is trained for each level (wind farm, bundle, total) of the hierarchy. Right: consistent forecasts are obtained via reconciliation.
    ]{%
        \includegraphics[width=\columnwidth]{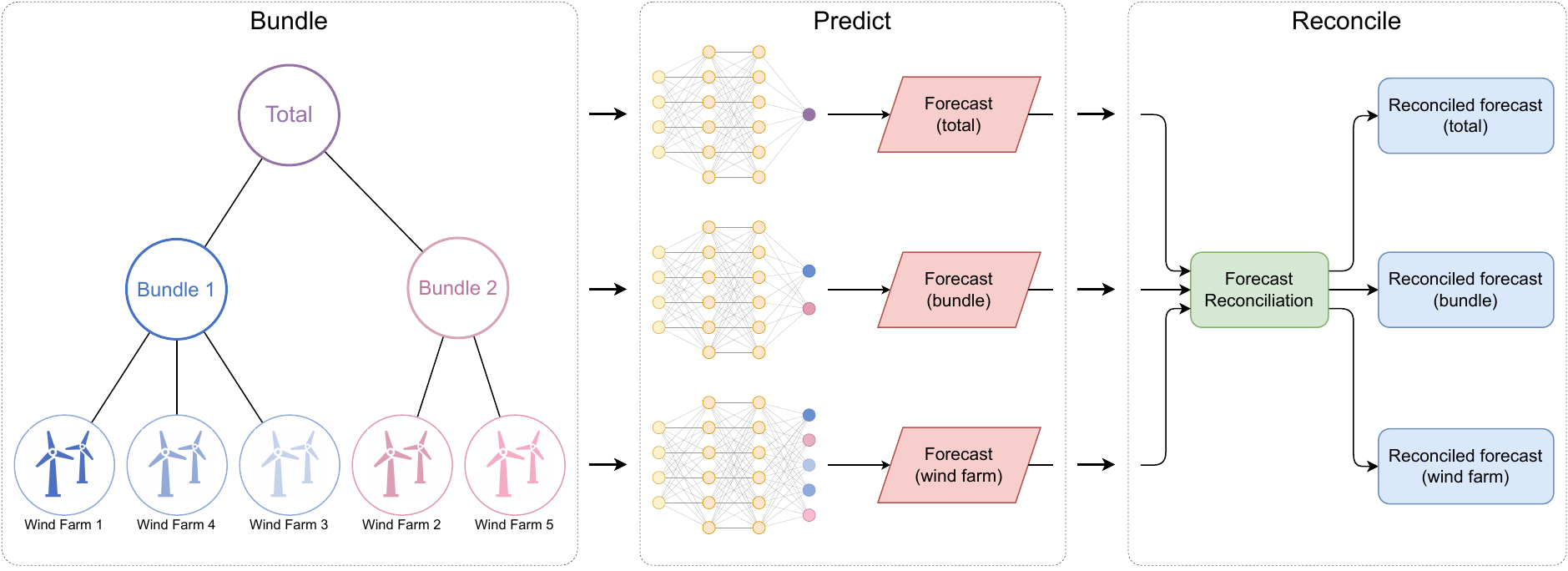}
        \label{fig:overview:proposed}
    }
    \caption{Illustration of the BPR Framework.}
    \label{fig:overview}
\end{figure}

Figure \ref{fig:overview} sketches the BPR framework.  A vanilla
approach to wind power forecasting is illustrated in Figure
\ref{fig:overview:vanilla}: it trains a model to predict
generator-level and total-level generations, resulting in forecasts
that may not be consistent. BPR is depicted in Figure
\ref{fig:overview:proposed}.  First, wind farms are grouped into $K$
bundles based on spatial and temporal features (see Section
\ref{sec:bundling}). Second, machine-learning models are trained to
deliver forecasts at the generator, bundle, and fleet levels.  Third,
a hierarchical reconciliation step produces consistent
forecasts.
\section{Asset Bundling}
\label{sec:bundling}

This section describes the asset-bundling methodology of BPR, from its
problem formulation to its solution algorithm.  The presentation
assumes a set of $N$ wind farms with given location, nameplate
capacity, and historical output $\x_{\mathcal{T}}$ where $\mathcal{T}\,
{=} \, \{ 1, ..., T \}$.  The distance between wind farms $i$ and $j$
is denoted by $D_{ij}$.

\subsection{Problem Formulation}
\label{sec:bundling:formulation}

The asset bundling problem consists in grouping $N$ wind farms into $K
\, {\in} \, \{1, .., N\}$ groups, referred to as \emph{bundles}, so as
to optimize a certain criterion.  Setting $K \, {=} \, 1$ aggregates all
wind farms into a single bundle, whereas $K \, {=} \, N$ considers all
wind farms individually. The asset-bundling problem can be modeled as
a combinatorial optimization problem.

Model \ref{model:bundling} presents an optimization model for the
asset-bundling problem and a criterion $f$. Binary variable
$\lambda_{k, i}$ takes value $1$ iff wind farm $i$ belongs to bundle
$k$. Constraints \eqref{eq:bundling:bundle_time_series} capture the
bundled time series $\z_{\mathcal{T}} \, {\in} \, \mathbb{R}^{K \times
  T}$.  Namely, the output of bundle $k$ at time $t$ is $\z_{k, t} \,
{=} \, \sum_{i} \lambda_{k, i} \x_{i, t}$, i.e., it is the combined
output of all wind farms assigned to bundle $k$.  Constraints
\eqref{eq:bundling:non_empty_bundle} and
\eqref{eq:bundling:set_partitioning} ensure that each bundle is
non-empty and that each wind farm is assigned to exactly one bundle.
Constraints \eqref{eq:bundling:distance_cutoff} enforce that two wind
farms cannot be assigned to the same bundle if the distance between
them exceeds a maximum diameter $\bar{D}$.  Finally, the objective
\eqref{eq:bundling:objective} captures properties of the bundled time
series $\z_{\mathcal{T}}$ and of the bundling assignment $\Lambda$;
possible choices for the objective function $f$ are described next.

    \begin{model}[!t]
        \caption{The Asset-Bundling Problem}
        \label{model:bundling}
        \small
        \textbf{Input:} Time series $\x_{\mathcal{T}}$, distance matrix $D$, distance cutoff $\bar{D}$\\
        \textbf{Variables:} $\lambda_{k, i}$ is $1$ if wind farm $i$ is in bundle $k$, $0$ otherwise
        \begin{subequations}
        \label{eq:bundling}
        \begin{align}
            \min_{\lambda} \quad & f(\z_{\mathcal{T}}, \Lambda) \label{eq:bundling:objective}\\
            \text{s.t.} \quad
                & \label{eq:bundling:bundle_time_series}
                    \z_{\mathcal{T}} = \Lambda \x_{\mathcal{T}}\\
                & \label{eq:bundling:non_empty_bundle}
                    \sum_{i \in \mathcal{N}} \lambda_{k, i} \geq 1
                    && \forall k \in \mathcal{K}\\
                & \sum_{k=1}^{K} \lambda_{k, i} = 1
                    && \forall i \in \mathcal{N}
                    \label{eq:bundling:set_partitioning}\\
                & D_{ij}(\lambda_{k, i} + \lambda_{k, j}) \leq 2 \bar{D}
                    && \forall (i, j) \in \mathcal{N}, k \in \mathcal{K}
                    \label{eq:bundling:distance_cutoff}
                    \\
                & \Lambda \in \{0, 1\}^{K \times N}
                    \label{eq:bundling:binary}
        \end{align}
        \end{subequations}
    \end{model}
    
\subsection{Bundling Criteria}
\label{sec:bundling:criteria}

This section describes various bundling criteria, which capture
desirable properties of the bundled time series.

\subsubsection{Variance}
\label{sec:bundling:criteria:Var}
    
\begin{figure}[!t]
\centering
\includegraphics[height=0.43\columnwidth]{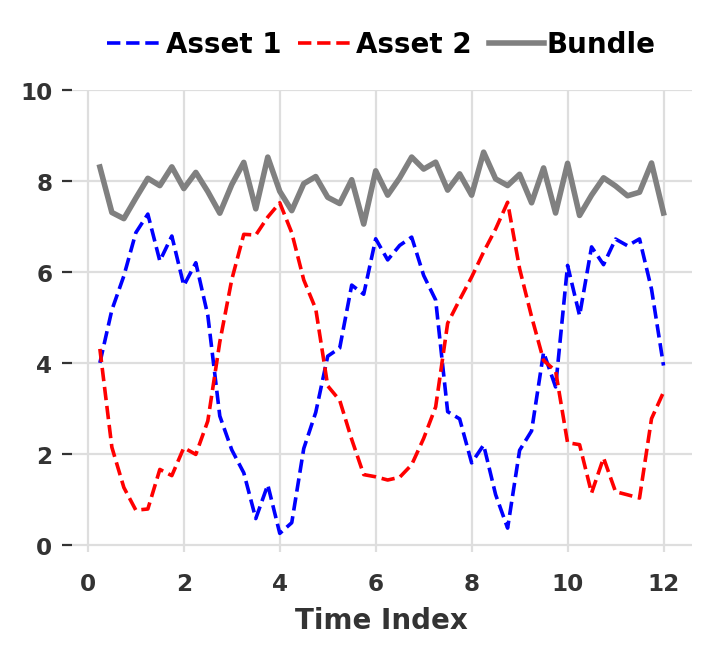}
\hfill
\includegraphics[height=0.43\columnwidth]{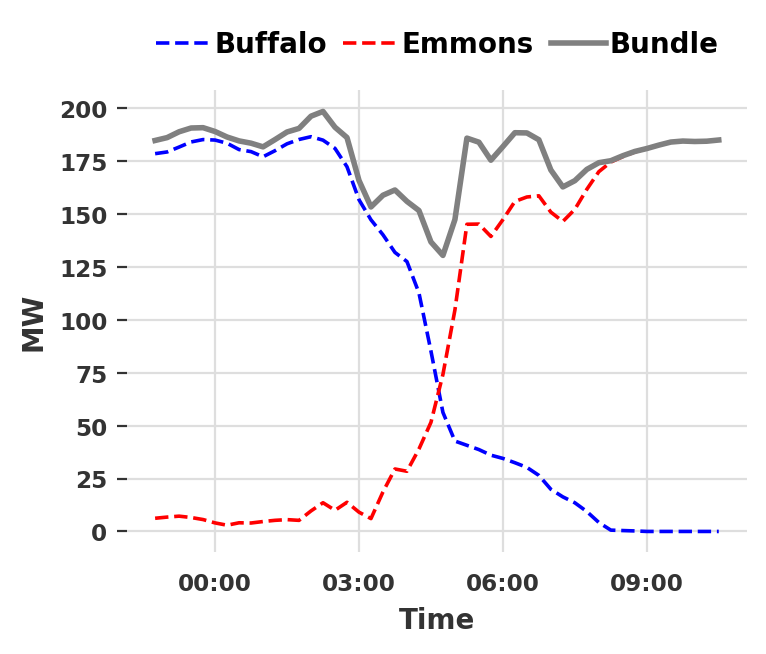}
\caption{Illustration of variance-based bundling. Left: synthetic example. Right: real example with two MISO wind farms from \cite{nrelperformdata}. In each case, the two assets are negatively correlated, and bundling yields a reduction in variance.}
\label{fig:bundling:illustration_savar}
\end{figure}

A natural bundling criterion is to minimize the total variance of the
bundles, i.e., 

\begin{align}
            \label{eq:bundling:variance:obj}
            f(\z_{\mathcal{T}}, \Lambda) &=
                \sum_{k=1}^{K} \Var(\z_{k, \mathcal{T}}) 
                = \tr(\Lambda \mathbf{\Sigma} \Lambda^{\top}),
\end{align}
where $\mathbf{\Sigma} \, {\in} \, \mathbb{R}^{N \times N}$ is the
empirical covariance matrix of $\x_{\mathcal{T}}$, $\tr$ stands for the trace of a matrix, and $\top$ represents the matrix (or vector) transpose.  Variance-based
bundling is illustrated in Figure
\ref{fig:bundling:illustration_savar}.  Note that bundling reduces the
total variance only if negatively correlated assets are bundled
together.  Therefore, a key factor for its success is the proportion
of negative correlations among the original assets.

\subsubsection{Seasonal-Adjusted Variance}
\label{sec:bundling:criteria:SAVar}

A large proportion of wind farms may exhibit positive correlations
over a given study period due to, e.g., common daily patterns.  This
is detrimental to variance-based bundling, which relies on negative
correlations. To mitigate this effect, the seasonal-adjusted variance
is defined as

\begin{align}
            \label{eq:bundling:SAVar}
            \SAVar(\x_{\mathcal{T}}) = \sum_{i \in \mathcal{N}} \Var(\x_{i, \mathcal{T}} - \boldsymbol{\mu}_{\mathcal{T}}),
        \end{align}
        where the average $\boldsymbol{\mu}_{\mathcal{T}} \, {=} \, \frac{1}{N} \sum_{i \in \mathcal{N}} \x_{i, \mathcal{T}}$ captures the seasonal trend.
        This yields the objective function
        \begin{align}
            \label{eq:bundling:SAVar:obj}
            f(\z_{\mathcal{T}}, \Lambda) 
            &= 
            \SAVar(\z_{\mathcal{T}})
            =\tr(\Lambda \mathbf{\tilde{\Sigma}} \Lambda^{\top}),
\end{align}

\noindent
where $\mathbf{\tilde{\Sigma}}$ is the covariance matrix of the
seasonal-adjusted time series $\mathbf{\tilde{x}}_{\mathcal{T}} \, {=}
\, \x_{\mathcal{T}} {-} \boldsymbol{\mu}_{\mathcal{T}}$.
Seasonal-adjusted variance-based bundling promotes bundles that behave
similarly to the mean of the assets, which is expected to improve the
quality of the forecast when the mean exhibits simple patterns.

\subsubsection{Intermittency Index}
\label{sec:bundling:criteria:Imcy}
    
\begin{figure}[!t]
        \centering
        \includegraphics[height=0.43\columnwidth]{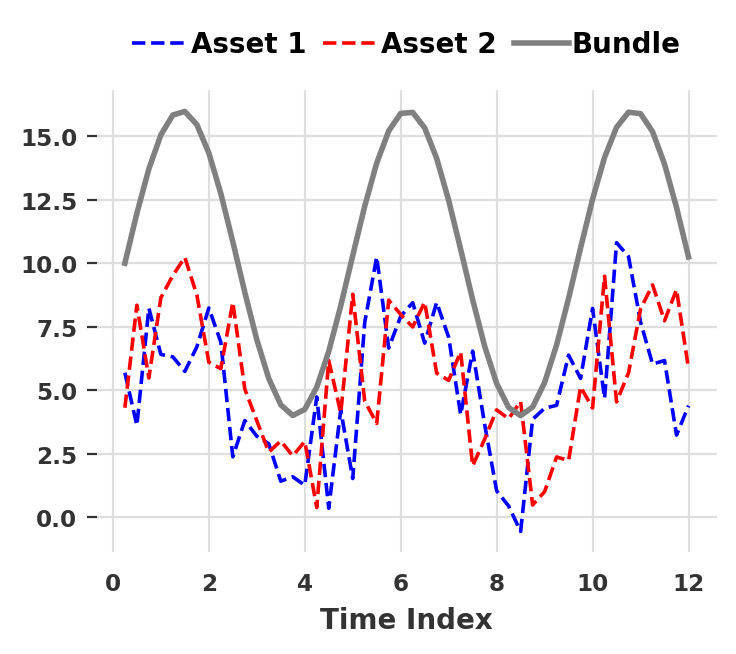}
        \hfill
        \includegraphics[height=0.43\columnwidth]{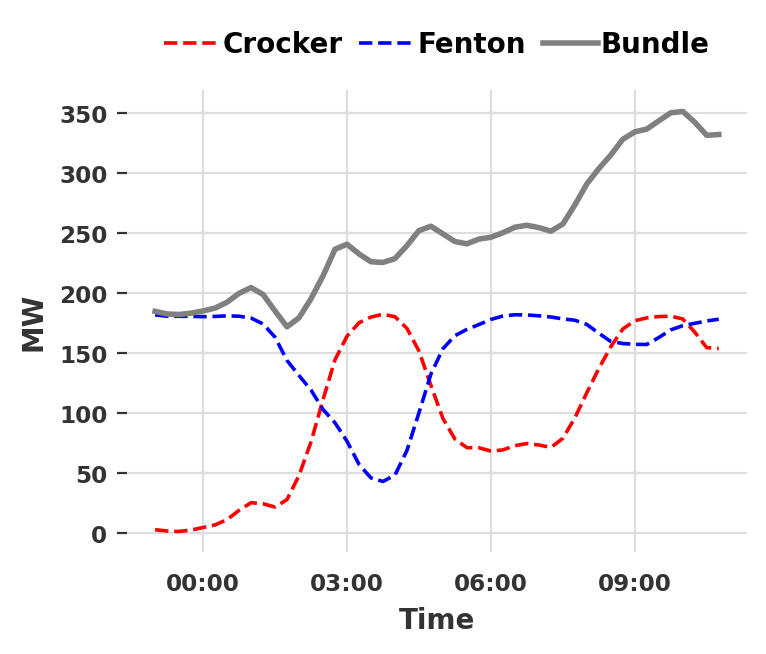}
        \caption{Illustration of intermittency index-based bundling. Left: synthetic example. Right: real example with two MISO wind farms from \cite{nrelperformdata}.
        In each case, bundling yields a reduction in intermittency: the bundled time series exhibits fewer high-frequency oscillations and clearer long-term patterns.}
        \label{fig:bundling:illustration_imcy}
\end{figure}

The \emph{intermittency index} of $\x_{\mathcal{T}}$ is
\begin{align}
\label{eq:bundling:intermittency}
\Intermittency(\x_{\mathcal{T}}) = \sum_{i \in \mathcal{N}} \Var(\xdot_{i, \mathcal{T}}),
\end{align}
 where $\xdot_{i, t} \, {=} \, \x_{i, t} {-} \x_{i, t-1}$; $\zdot$ is
 defined similarly.  Substituting $\zdot \, {=} \, \Lambda \xdot$, the
 objective function \eqref{eq:bundling:objective} becomes
\begin{align}
            \label{eq:bundling:intermittency:obj}
            f(\z_{\mathcal{T}}, \Lambda) = \Intermittency(\z_{\mathcal{T}}) = \tr(\Lambda \mathbf{\dot{\Sigma}} \Lambda^{\top}),
\end{align}

\noindent
where $\mathbf{\dot{\Sigma}}$ is the covariance matrix of
$\xdot_{\mathcal{T}}$.  Intermittency-based bundling is illustrated in
Figure \ref{fig:bundling:illustration_imcy} where the bundled time
series exhibit long-term patterns that are easier to learn. In the
context of power systems operations, the ability of
intermittency-based bundling to identify long-term trends allows to
better anticipate and manage ramping needs.

\subsection{The Bundling Algorithm}
\label{sec:bundling:algorithm}

The objective functions in Eqs. \eqref{eq:bundling:variance:obj},
\eqref{eq:bundling:SAVar:obj} and
\eqref{eq:bundling:intermittency:obj} share the same intrinsic
structure, differring only in the choice of the covariance matrix
$\mathbf{\Sigma}$, $\mathbf{\tilde{\Sigma}}$, or
$\mathbf{\dot{\Sigma}}$. As a result, for each proposed bundling
criterion, Problem \eqref{eq:bundling} is a convex mixed-integer
quadratic programming (MIQP) problem, which can be solved using MIQP
solvers like Gurobi or CPLEX.  However, for realistic test cases, the
asset-bundling problem is intractable for MIQP solvers, mainly due to
its intrinsic complexity and the symmetry of set-partitioning
constraints \eqref{eq:bundling:set_partitioning}.  The latter are
notoriously detrimental to the performance of MIQP solvers.

To remedy these scalability issues, the BPR framework uses a fast
greedy algorithm presented in Algorithm \ref{algo:bundling:greedy}.
For simplicity, Algorithm \ref{algo:bundling:greedy} is presented for
the case of Variance-based bundling. The algorithm takes as input the
time series $\x$ and the number of bundles $K$, and initializes a set
of $N$ bundles, each corresponding to an individual wind farm.  Then,
at each step, the algorithm aggregates the two most negatively
correlated bundles, subject to the maximum diameter constraint.
Algorithm \ref{algo:bundling:greedy} is computationally fast, and was
found to produce close-to-optimal solutions in early experiments.

    \begin{algorithm}[!t]
        \caption{The Greedy Asset-Bundling Algorithm.}
        \label{algo:bundling:greedy}
        \SetAlgoLined
        \textbf{Input:} Time series $\x_{\mathcal{T}}$, distance matrix $D$, maximum diameter $\bar{D}$, number of bundles $K$\\
        \textbf{Initialization:} $\mathcal{K} \leftarrow \{ \{1\}, ..., \{N\} \}$\\ 
        \While{$|\mathcal{K}|> K$}{
            $\z_{k} \gets \sum_{i \in \mathcal{K}_{k}} \x_{i, \mathcal{T}},$
                \hfill $k = 1, ..., |\mathcal{K}|$\\
            $\tilde{D}_{k, l} \gets \max \{ D_{ij} | (i, j) {\in} \mathcal{K}_{k} {\times} \mathcal{K}_{l}\},$
                \hfill $k, l = 1, ..., |\mathcal{K}|$\\
            $(k, l) \gets \arg \min \left\{
                \text{Cov} \left(\z_{k}, \z_{l} \right)
                \, \middle| \, 
                \tilde{D}_{kl} \leq \bar{D}
            \right\}$\\
            $\mathcal{K} \gets (\mathcal{K} \setminus \{\mathcal{K}_{k}, \mathcal{K}_{l}\}) \cup (\mathcal{K}_{k} \cup \mathcal{K}_{l})$
        }
        \textbf{Return:} set of bundles $\mathcal{K}$
    \end{algorithm}

\section{Forecasting}
\label{sec:forecasting}

\newcommand{\X}{\mathbf{{X}}}
\newcommand{\Xhat}{\mathbf{\hat{X}}}
\newcommand{\hhat}{\mathbf{\hat{h}}}
\newcommand{\hrec}{\mathbf{\tilde{h}}}
\newcommand{\zrec}{\mathbf{\tilde{z}}}
\newcommand{\Xrec}{\mathbf{\tilde{X}}}
\newcommand{\xrec}{\mathbf{\tilde{x}}}

This section presents the hierarchical forecasting and reconciliation
component of BPR, the architecture of the ML models considered in the
experiments, and the metrics for evaluating the quality of the
forecasts.  The section assumes a fixed bundling of $N$ wind farms
into $K$ bundles, denoted by $\Lambda \, {\in} \, \{0, 1 \}^{K \times
  N}$.  For $t \, {\in} \, \mathbb{Z}$, $\z_{t} \, {=} \, \Lambda
\x_{t} \, {\in} \, \mathbb{R}^{K}$ and $\X_{t} \, {=} \, e^{\top}
\x_{t} \, {\in} \, \mathbb{R}$ denote the bundle-level and total-level
wind power output, respectively.  The forecast horizon at time $t$ is
$\T_{t} \, {=} \, \{t{+}1, ..., t{+}T\}$.

\subsection{Reconciliation}
\label{sec:forecast:hierarchical}

BPR works on a three-level hierarchical structure illustrated in
Figure \ref{fig:overview:proposed}: generator-level ($\x$),
bundle-level ($\z)$, and total-level ($\X$).  The paper leverages this
structure by training three models that predict each level of this
hierarchy: at time $t$, three forecasts $\xhat_{\T_{t}},
\zhat_{\T_{t}}, \Xhat_{\T_{t}}$ are produced that predict the future
generator-level, bundle-level, and total-level time series
$\x_{\T_{t}}, \z_{\T_{t}}, \X_{\T_{t}}$, respectively.
    
The reconciliation aims at making these predictions consistent. Given
forecasts $\xhat, \zhat, \Xhat$ at the asset, bundle, and fleet
levels, reconciliation produces new forecasts $\xrec, \zrec, \Xrec$
such that
\begin{align*}
        \forall k, \zrec_{k, \T_{t}} &= \sum_{i \in \N} \lambda_{k, i} \xrec_{i, \T_{t}}\\
        \Xrec_{\T} &= \sum_{i \in \N} \xrec_{i, \T} = \sum_{k} \zrec_{k, \T}
\end{align*}

\noindent
More formally, forecast reconciliation \cite{Hyndman2011_ForecastHierarchicalTimeSeries} takes as
input incoherent forecasts, and produces the most accurate consistent
forecasts.  Let $\hhat = (\Xhat, \zhat, \xhat)$ and $S$ be the summing
matrix
\begin{align}
S &= \begin{bmatrix} e & \Lambda^{\top} & I \end{bmatrix}^{\top} \in \mathbb{R}^{(N+K+1)\times N}.
\end{align}
Reconciliation searches for a matrix $G \, {\in} \, \mathbb{R}^{N\times(N+K+1)}$ such that
$\hrec_{\tau} = S G \hhat_{\tau}$.
In addition, to improve accuracy further, BPR computes $T$
matrices $G_{1}, ..., G_{T}$ such that
\begin{align*}
\forall \tau \in \{1, ..., T\}, \ \hrec_{\tau} &= S G_{\tau} \hhat_{\tau}
\end{align*}
to capture the variances at different lead times $\tau$. BPR uses the
MinT optimal reconciliation approach from \cite{minT2019}, with a
weighted least squares (WLS) estimator using variance scaling.
Thereby, each $G_{\tau}$ has the form
\begin{align}
G_{\tau} &= (S^{\top} W_{\tau}^{-1} S)^{-1} S^{\top} W_{\tau}^{-1}
\end{align}
where $W_{\tau} = \text{Diag}(\hat{W}_{\tau})$ and
\begin{align}
\hat{W}_{\tau} &= \frac{1}{M} \sum_{t \in \mathcal{M}} \epsilon_{t, \tau} \epsilon_{t, \tau}^{\top}
\end{align}
\noindent
and $\epsilon_{t, \tau} = (\hhat_{t, \tau} - \mathbf{h}_{t, \tau})$.
For the case at hand, WLS reconciliation achieved better performance
than other MinT-based reconciliation approaches, such as ordinary
least squares, WLS using structural scaling, and MinT-cov
\cite{minT2019}.

\subsection{Forecasting Models}
\label{sec:forecasting:architectures}

\newcommand{\Persistence}{\texttt{Per}}
\newcommand{\Ridge}{\texttt{Lin}}
\newcommand{\LSTM}{\texttt{RNN}}
\newcommand{\Transformer}{\texttt{Tfm}}
\newcommand{\TCN}{\texttt{TCN}}
\newcommand{\TFT}{\texttt{TFT}}

BPR was instantiated with various learning architectures for the
experimental evaluation.  These are summarized in Table
\ref{tab:forecasting:architectures}, which reports, for each
learning model, whether it supports past, future, and static
covariates (see \cite{Herzen2022_Darts}).  The co-variates may include
static information about each wind farm (e.g., its location and
model), temporal information such as hour of the day or month of the
year, and past and forecasted weather-related information such as
temperature, humidity, wind speed and direction.  Covariates have been
shown to improve forecast accuracy
\cite{EIKELAND2022100239,cao2012forecasting}.

    \begin{table}[!t]
        \centering
        \caption{Overview of ML Architectures.}
        \label{tab:forecasting:architectures}
        \begin{tabular}{ccl}
            \toprule
            Model & Covariates$^{*}$ & Description \\
            \midrule
            \Persistence
                & \xmark \, \xmark \, \xmark
                & Persistence model\\
            \Ridge
                & \cmark \, \cmark \, \cmark
                & Linear model\\
            \LSTM 
                & \xmark \, \cmark \, \xmark 
                & Recurrent Neural Network\\
            \Transformer 
                & \cmark \, \xmark \, \xmark 
                & Transformer\\
            \TCN
                & \cmark \, \xmark \, \xmark
                & Temporal Convolution Transformer\\
            \TFT 
                & \cmark \, \cmark \, \cmark
                & Temporal Fusion Transformer\\
            \bottomrule
        \end{tabular}\\
        {\footnotesize$^{*}$Whether a model supports past/future/static covariates \cite{Herzen2022_Darts}}
    \end{table}
    
    \subsubsection{Persistence Model}
        The Persistence model (\Persistence) is a naive baseline whose
        prediction for the entire forecasting horizon is equal to the
        last observed value.

    \subsubsection{Linear Model}
        The linear regression model is trained using ridge regression.
        Without loss of generality, a separate model is trained for
        each bundle, which makes training more efficient.

    \subsubsection{Recurrent Neural Network}
        The Recurrent Neural Network (RNN) architecture follows the
        state-of-the-art DeepAR model \cite{Salinas2020_DeepAR}.  It
        is applied in an auto-regressive way to compute multi-step
        predictions.  The RNN models are trained with mean squared
        error (MSE) loss.
        
    \subsubsection{Transformer models}
        The baseline transformer model follows the approach of
        \cite{NIPS2017_3f5ee243}.  Namely, it employs a multi-head
        attention mechanism inside an encoder-decoder,
        sequence-to-sequence architecture.

    \subsubsection{Temporal Convolution Transformer}
        The Temporal Convolution Transformer (TCN) architecture
        follows the model from \cite{TCNPaper}.  Unlike the baseline
        transformer model, TCN uses stacked dilated causal
        convolutional neural networks to extract features from
        historical observations.

    \subsubsection{Temporal Fusion Transformer}
        The Temporal Fusion Transformer (TFT) architecture
        \cite{lim2019temporal} is a deep learning architecture
        designed for multi-horizon forecasting tasks with a mixed-type
        inputs, e.g., past observations and past, future, and static
        covariates.  This allows the integration of weather-based,
        wind speed forecasts into the model, which improves
        longer-term predictions.  The TFT architecture uses a
        combination of recurrent layers and self-attention layers to
        learn temporal relationships at different scales and provide
        interpretable insights into the temporal dynamics.  The model
        utilizes specialized components to select relevant features,
        and gating layers to suppress unnecessary components.
        
\subsection{Evaluation Metrics}
\label{sec:forecasting:metrics}

    \newcommand{\MAE}{\text{MAE}}
    \newcommand{\NMAE}{\text{NMAE}}
    \newcommand{\RMSE}{\text{RMSE}}
    \newcommand{\VS}{\text{VS}}
    \newcommand{\ED}{\text{ED}}
    
    The experimental evaluation uses various metrics to evaluate the
    quality of wind power forecasts.  Each metric is presented for
    asset-level times series $\x$ in what follows, but they are
    similar for the bundle and fleet levels.  See
    \cite{Messner202_EvaluationWindForecast} for a detailed discussion
    of wind power forecast evaluation.

    \subsubsection{Normalized Mean Absolute Error}
        
        The Normalized Mean Absolute Error (NMAE) is defined as
        \begin{align}
            \label{eq:forecasting:metrics:NMAE}
            \NMAE = \frac{1}{MNT} \sum_{t \in \mathcal{M}} \sum_{i \in \N} \frac{\| \mathbf{{x}}_{i, \T_{t}} - \mathbf{\hat{x}}_{i, \T_{t}} \|_{1}}{\mathbf{\bar{x}}_{i}} 
        \end{align}
        where $\mathbf{\bar{x}}_{i}$ is the maximum capacity of wind farm $i$ and $\|\cdot\|_{1}$ is the vector $\ell_1$ norm.
    \subsubsection{Root-mean-square Error}
        The Root-mean-square Error (RMSE) is expressed as 
        \begin{align}
            \label{eq:forecasting:metrics:RMSE}
            \RMSE = \sqrt{\frac{1}{MTN}  \sum_{i \in \N} \sum_{t \in \mathcal{M}} \left\| \mathbf{x}_{\T_{t}} - \mathbf{\hat{x}}_{\T_{t}} \right\|_{2}^2}
        \end{align}
        where $\|\cdot\|_{2}$ is the vector $\ell_2$ norm.
        $\RMSE$ better captures large prediction errors than $\NMAE$.
    \subsubsection{Variogram Score}
        The Variogram Score (\VS) of order $p > 0$ is given by
        \begin{align}
            \label{eq:forecasting:metrics:VS}
            \VS_{p} =
            \frac{1}{M} \sum_{t \in \mathcal{M}} \sum_{i, j \in \mathcal{N}} \sum_{\tau, \tau' \in \T_{t}} \left( 
                \mathbf{\Delta x}^{p}_{i, j, \tau, \tau'} -  \mathbf{\Delta \hat{x}}^{p}_{i, j, \tau, \tau'}
            \right)^{2}
        \end{align}
        where $\mathbf{\Delta x}^{p}_{i, j, \tau, \tau'}$ and $\mathbf{\Delta \hat{x}}^{p}_{i, j, \tau, \tau'}$ are defined as
        \begin{align*}
            \mathbf{\Delta x}^{p}_{i, j, \tau, \tau'}
                &= \left| \mathbf{x}_{i, \tau} - \mathbf{x}_{j, \tau'} \right|^{p}, &
            \mathbf{\Delta \hat{x}}^{p}_{i, j, \tau, \tau'}
                &= \left| \mathbf{\hat{x}}_{i, \tau} - \mathbf{\hat{x}}_{j, \tau'} \right|^{p}
            .
        \end{align*}
        The paper uses $p \, {=} \, {1}/{2}$, as recommended in \cite{scheuerer_hamill_2015}.
        Note that computing $\VS_{p}$ requires $\mathcal{O}(MN^{2}T^{2})$ operations, which is computationally expensive for the case at hand.
        Therefore, the paper only reports \VS{} for total wind predictions, which reduces the complexity to $\mathcal{O}(MT^{2})$.
    
    \subsubsection{Energy Distance}
        The Energy Distance (\ED) is
        \begin{align}
            \label{eq:forecasting:metrics:ED}
            \ED &= \frac{2}{M} \sum_{t \in \mathcal{M}} \left\| \mathbf{x}_{\T_{t}} - \mathbf{\hat{x}}_{\T_{t}} \right\|_{F}
        \end{align}
        where $\| \cdot \|_{F}$ denotes the Frobenius norm.

\section{Experimental Evaluation}
\label{sec:experiments}

\subsection{Data}

The results are presented on the NREL dataset \cite{nrelperformdata}
that contains 283 time series of existing wind farms in MISO system
for 2018 and 2019 at different time granularities (1 hour and 15
minutes). This dataset is complemented by a separate dataset of
weather forecasts from the European Center for Medium-Range Weather
Forecasts (ECMWF) \cite{ecmwf}. These forecasts are collected on a
1-degree by 1-degree grid for the MISO region and used as covariates
in the learning models. They are generated twice a day, at noon and
midnight in UTC time, and have a 6-hour granularity.

The weather forecasts time series are
interpolated from 6-hour granularity down to the desired granularity
of the forecasts.

\subsection{Experimental Setting}

The experiments consider both short-term and day-ahead forecasting
tasks. Short-term forecasts predict wind power generation over
the next 6 hours with a granularity of 15 minutes; it is the
forecasting task needed for the Look-Ahead Commitment (LAC) in the
MISO market clearing pipeline.
Short-term forecasts are purely data-driven, because the original ECMWF weather forecasts have a granularity of 6 hours and are produced only twice a day.
Day-ahead forecasts predict wind power generation over the next 48 hours with a granularity of 1 hour; it is the
forecasting task needed for the day-ahead reliability commitment
(FRAC) in the MISO market clearing pipeline.
The day-ahead forecasting task is data-driven as well but it also uses co-variates given by the latest available ECMWF weather forecasts.

The evaluation uses a test set comprising
five weeks in 2019, i.e., the weeks of January 8th, March 8th, May
8th, July 8th, and September 8th.
For short-term predictions, the training dataset consists of 90
days of historical data immediately preceding each test week. 
The training for day-ahead forecasting uses a more extensive dataset
spanning a year. The evaluation uses two bundle sizes: 10 and 50
bundles. The diameter constraints are set to 500 kilometers for 50
bundles and 800 kilometers for 10 bundles. The evaluation trains three
types of models for (1) fleet level ($K = 1$), (2) bundle level ($K =
10$ and $K = 50$), and (3) asset level ($K = 283$).

\begin{figure}[!t]
    \centering
    \includegraphics[width=0.48\columnwidth]{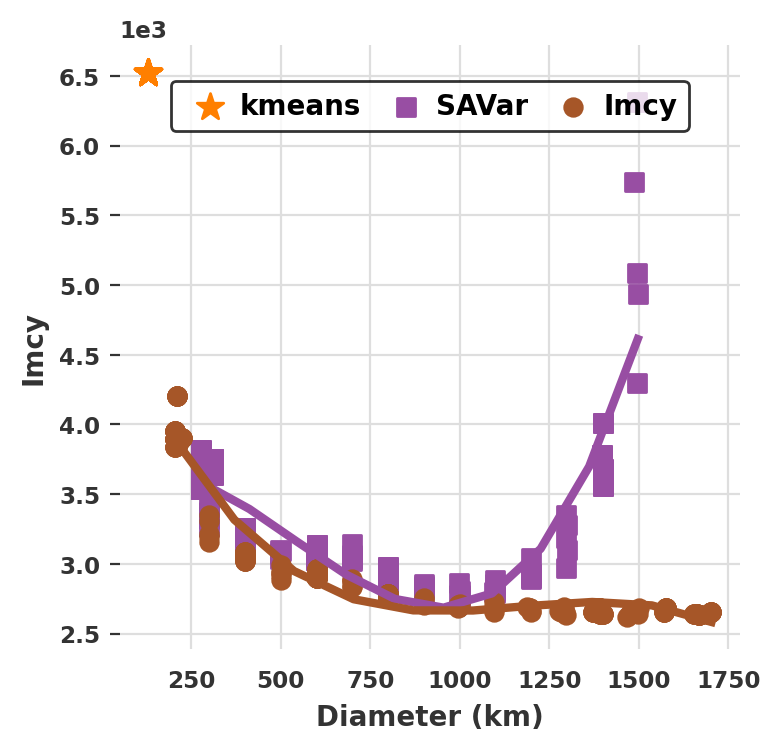}
    \hfill
    \includegraphics[width=0.48\columnwidth]{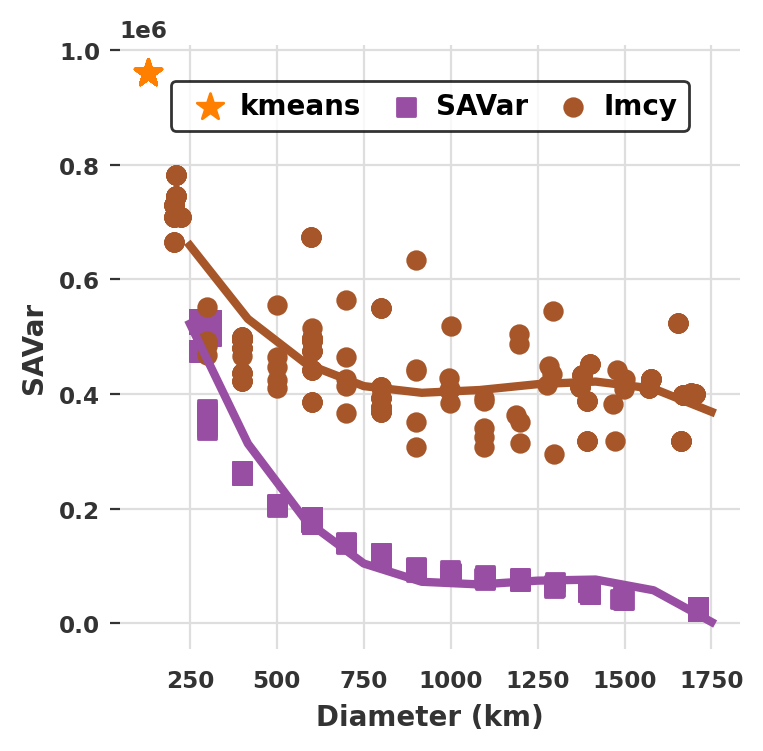}\\
    \caption{Sensitivity of Imcy and SAVar with respect to diameter for three methods when considering 50 bundles. Bundle samples are generated by adjusting diameter constraints. A cubic trend line is fitted for better visualization. }
    \label{fig: imcy/savar v.s. diameter}
\end{figure}

Figure \ref{fig: imcy/savar v.s. diameter} depicts the tradeoffs
between the diameter constraints for the bundles and the quality of
the bundling criteria (Imcy and SAVar). It shows that it is possible
to obtain low values for the intermittency and the SAVar criteria with
a reasonable diameter constraint. This informed the selection of the
diameter constraints specified earlier.

\subsection{Prediction Performance}
\label{sec:experiment:prediction}

Tables \ref{tab:results:ST-rcc} and \ref{tab:results:DA-rcc} present the results obtained by BPR for the short-term and day-ahead forecasting tasks, respectively.
All results are averaged over the five selected weeks.
For each model architecture, each table reports: the number of bundles (K), the bundling criterion, and performance metrics of fleet-level forecasts (NMAE, RMSE, ED, VS) and asset-level forecasts (NMAE, RMSE, ED).
Rows with $K \, {=} \,1$ correspond to a baseline hierarchical forecast without any bundling, i.e., the corresponding forecasts are obtained by reconciling  an asset-level and a fleet-level forecasts.
A detailed analysis of each forecasting task is presented next.

\paragraph{Short-Term Predictions}

\begin{table}[!t]
    \centering
    \caption{BPR for 6 Hours Forecasting: Accuracy Results.}
    \label{tab:results:ST-rcc}
    \resizebox{\columnwidth}{!}{
    \begin{tabular}{crrrrrrrrrr}
        \toprule
        && & \multicolumn{4}{c}{Fleet} & \multicolumn{3}{c}{Asset}\\
        \cmidrule(lr){4-7} \cmidrule(lr){8-10}
        Model 
            & K
            & criterion
            & \multicolumn{1}{c}{NMAE}
            & \multicolumn{1}{c}{RMSE}
            & \multicolumn{1}{c}{ED}
            & \multicolumn{1}{c}{VS}
            & \multicolumn{1}{c}{NMAE}
            & \multicolumn{1}{c}{RMSE}
            & \multicolumn{1}{c}{ED}
            \\
        \midrule
        \Persistence 
            & --  &--  
                &7.6 &2444 &18797 &631229 
                &13.9 &25.3 &3911 \\ 
        \midrule
        \Ridge 
            &\textbf{1} &-- 
                &6.5 &2038 &16370 &171212 
                &14.0 &22.4 &3524 \\ 
            \cmidrule{2-10}        
            \textbf{} &\textbf{10} &Imcy 
                &6.2 &1958 &15637 &165624 
                &13.9 &22.0 &3469 \\
            \textbf{} &\textbf{} &SAVar 
                &6.1 &1955 &15631 &167449 
                &13.9 &22.0 &3471 \\
            \textbf{} &\textbf{} &kmeans 
                &6.2 &1959 &15662 &166057 
                &13.9 &22.1 &3471 \\
            \cmidrule{2-10}
            \textbf{} &\textbf{50} &Imcy 
                &6.0 &1925 &15200 &163855 
                &13.7 &21.3 &3355 \\
            \textbf{} &\textbf{} &SAVar 
                &6.1 &1940 &15407 &163822 
                &13.8 &21.9 &3437 \\
            \textbf{} &\textbf{} &kmeans 
                &6.0 &1930 &15348 &165105 
                &13.7 &21.7 &3406 \\
        \midrule
        \LSTM 
            &\textbf{1} &-- 
                &8.1 &2524 &20275 &199126 
                &15.6 &24.8 &3928 \\ 
            \cmidrule{2-10}
            \textbf{} &\textbf{10} &Imcy 
                &6.7 &2114 &16927 &175617 
                &15.2 &23.5 &3752 \\
            \textbf{} &\textbf{} &SAVar 
                &7.4 &2284 &18374 &191918 
                &15.4 &23.9 &3807 \\
            \textbf{} &\textbf{} &kmeans 
                &8.1 &2463 &20057 &196443 
                &15.6 &24.6 &3904 \\
            \cmidrule{2-10}
            \textbf{} &\textbf{50} &Imcy 
                &8.6 &2549 &20919 &199591 
                &15.6 &24.4 &3863 \\
            \textbf{} &\textbf{} &SAVar 
                &7.1 &2199 &17534 &175830 
                &15.2 &23.8 &3781 \\
            \textbf{} &\textbf{} &kmeans 
                &7.7 &2387 &19316 &202821 
                &15.3 &24.0 &3807 \\ 

        \midrule
        \Transformer 
            &\textbf{1} &-- &7.2 &2054 &17026 &173000 &15.3 &22.7 &3615 \\ 
            \cmidrule{2-10}
            \textbf{} &\textbf{10} &Imcy &6.4 &1878 &15348 &168573 &15.2 &22.3 &3554 \\
            \textbf{} &\textbf{} &SAVar &6.5 &1893 &15508 &167283 &15.2 &22.3 &3559 \\
            \textbf{} &\textbf{} &kmeans &6.8 &1956 &16121 &170853 &15.3 &22.5 &3591 \\
            \cmidrule{2-10}
            \textbf{} &\textbf{50} &Imcy &6.5 &1885 &15517 &170166 &15.2 &22.3 &3547 \\
            \textbf{} &\textbf{} &SAVar &6.7 &1931 &15775 &161601 &15.2 &22.4 &3561 \\
            \textbf{} &\textbf{} &kmeans &6.8 &1950 &16029 &162314 &15.2 &22.5 &3572 \\ 

        \midrule
        \TCN 
            &\textbf{1} &-- &10.6 &2811 &24388 &211609 &19.3 &27.7 &4428 \\ 
            \cmidrule{2-10}
            \textbf{} &\textbf{10} &Imcy &10.3 &2725 &23763 &202882 &19.3 &27.4 &4390 \\
            \textbf{} &\textbf{} &SAVar &10.4 &2738 &23851 &209404 &19.3 &27.5 &4405 \\
            \textbf{} &\textbf{} &kmeans &10.3 &2730 &23682 &209485 &19.3 &27.4 &4397 \\
            \cmidrule{2-10}
            \textbf{} &\textbf{50} &Imcy &10.0 &2672 &23182 &213046 &19.1 &27.0 &4328 \\
            \textbf{} &\textbf{} &SAVar &10.1 &2691 &23375 &207217 &19.1 &27.3 &4365 \\
            \textbf{} &\textbf{} &kmeans &10.2 &2735 &23552 &213180 &19.2 &27.3 &4374 \\ 
 
        \midrule
        \TFT &
            \textbf{1} &-- 
                &6.4 & 1899	& 15432 & 152863& 15.3 & 23.7 & 3755\\
            \cmidrule{2-10}
            \textbf{} &\textbf{10} &Imcy &5.8 &1753 &14163 &142315 &15.3 &23.5 &3723 \\
            \textbf{} &\textbf{} &SAVar &5.7 &1721 &13891 &142824 &15.3 &23.4 &3708 \\
            \textbf{} &\textbf{} &kmeans &6.0 &1780 &14387 &140360 &15.3 &23.5 &3725 \\
            \cmidrule{2-10}
            \textbf{} &\textbf{50} &Imcy &6.0 &1778 &14472 &145387 &15.3 &23.5 &3724 \\
            \textbf{} &\textbf{} &SAVar &6.1 &1797 &14689 &148588 &15.3 &23.5 &3729 \\
            \textbf{} &\textbf{} &kmeans &5.9 &1746 &14256 &142929 &15.2 &23.3 &3697 \\

        \midrule
        \TFT~+ \Ridge
            & \textbf{10} &  SAVar 
            & 5.8 & 1806 & 14574 & 152841 
            & 13.8 & 21.7 & 3423 \\ 
\bottomrule
\end{tabular}}
\end{table}

Observe first that, for short-term forecasting, the persistence model
emerges as a strong baseline, achieving a 7.6\% NMAE at the fleet
level, and a 13\% NMAE and a 25.2MW RMSE at the asset level. Only the
Ridge regression, transformer, and TFT models are capable of improving
over the baseline. For the TFT model with 10 SAVar bundles, BPR
achieves a NMAE of 5.7\% at the fleet level and a 15.3\% NMAE and a
23.5MW RMSE at the asset level.
The best results come from reconciling asset level forecasts from the Ridge regression and bundle level forecasts from TFT (\TFT+\Ridge, last row of the table).
The latter was found to produce the most accurate asset-level forecast, and the former to produce the most accurate bundle-level and fleet-level forecast.
In this case, BPR achieves a NMAE of 5.8\% at the
fleet level and a 18.8\% NMAE and a 21.7MW RMSE at the asset level.
When correlations are taken into account, as captured by ED and VS metrics, BPR achieves consistently better results, especially for VS at the fleet level.
{\em In summary, BPR improves the accuracy by about 25\% (NMAE, RMSE and ED) at the fleet level and by about 14\% (RMSE and ED) at the asset level over the baseline, and reduces VS by 75\% at the fleet level.}

When looking at a specific model (e.g., TFT), the main benefit of BPR
for short-term forecasting is at the fleet level. For instance, BPR on
TFT with SAVar and 10 bundles improves the fleet level accuracy by
about 10\% (NMAE, RMSE and ED) over the base setting ($K=1$); VS is also improved by 6\%.
The improvements of BPR at the asset level through
reconciliation is about 2\% (RMSE), again for TFT with SAVar and 10
bundles. Note also that the bundling criteria achieve rather similar
results: SAVar achieves consistently strong results overall, but there
is no clear overall winner. Similarly, there is no significant
difference between 10 and 50 bundles for short-term forecasting.

\paragraph{Day-Ahead Predictions}
The persistence model is not relevant for the day-ahead setting, 
so the discussion focuses on the benefits of BPR for specific models.
Nonetheless, the short-term forecasting results provide a good
reference point, indicating the high quality of the day-ahead
forecasts.

For day-ahead forecasting, the {\tt Imcy} criterion
dominates the others (almost always) and BPR with 50 bundles is
typically superior to BPR with 10 bundles.
In addition, BPR always improves against the baseline ($K{=}1$), especially when considering ED and VS metrics, which capture correlations.
BPR over the TFT model with 50 bundles and the {\tt Imcy} criterion achieves the best results overall.
{\em At the fleet level, BPR yields a 6--8\% improvement in NMAE and RMSE, and a 20\% improvement in ED and VS over the baseline.}
At the asset level, BPR yields a 3\% improvement in RMSE and ED.

\begin{table}[!t]
    \centering
    \caption{BPR for 48 hours Forecasting: Accuracy Results.}
    \label{tab:results:DA-rcc}
    \resizebox{\columnwidth}{!}{
    \begin{tabular}{crrrrrrrrrr}
        \toprule
        && & \multicolumn{4}{c}{Fleet} & \multicolumn{3}{c}{Asset}\\
        \cmidrule(lr){4-7} \cmidrule(lr){8-10}
        Model 
            & K
            & criterion
            & \multicolumn{1}{c}{NMAE}
            & \multicolumn{1}{c}{RMSE}
            & \multicolumn{1}{c}{ED}
            & \multicolumn{1}{c}{VS}
            & \multicolumn{1}{c}{NMAE}
            & \multicolumn{1}{c}{RMSE}
            & \multicolumn{1}{c}{ED}
            \\
            \midrule
        \LSTM 
            &\textbf{1} &-- &10.5	& 2848	& 38389 &	1370108	& 19.5	& 28.8	& 6673 \\
            \cmidrule{2-10}            
            &\textbf{10} &Imcy &9.8 &2655 &35722 &1228135 &19.3 &28.0 &6478 \\
            &\textbf{} &SAVar &9.8 &2662 &35818 &1242135 &19.3 &28.1 &6514 \\
            &\textbf{} &kmeans &10.2 &2743 &36973 &1283293 &19.3 &28.3 &6562 \\
            \cmidrule{2-10}
            &\textbf{50} &Imcy &9.2 &2444 &33102 &1167797 &18.9 &26.7 &6189 \\
            &\textbf{} &SAVar &10.3 &2749 &37135 &1296811 &19.2 &28.2 &6525 \\
            &\textbf{} &kmeans &10.1 &2720 &36611 &1331087 &19.1 &28.1 &6516 \\ 

    \midrule
        \TFT 
            &\textbf{1} &-- &8.8& 2480& 33236	& 1214637 & 	16.1 &	25.7& 5918 \\
            \cmidrule{2-10}
            &\textbf{10} &Imcy &8.4 &2364 &31586 &1117189 &16.0 &25.3 &5819 \\
            &\textbf{} &SAVar &8.6 &2391 &31915 &1139472 &16.0 &25.5 &5867 \\
            &\textbf{} &kmeans &8.6 &2408 &32227 &1153448 &16.0 &25.4 &5862 \\ 
            \cmidrule{2-10}
            &\textbf{50} &Imcy &8.1 &2324 &30835 &1099162 &15.9 &24.9 &5726 \\
            &\textbf{} &SAVar &8.3 &2330 &31178 &1119034 &15.9 &25.2 &5797 \\
            &\textbf{} &kmeans &8.2 &2302 &30615 &1092823 &15.8 &25.0 &5753 \\ 

    \bottomrule
    \end{tabular}}
\end{table}

\subsection{Comparison with NREL Forecasts}
\label{sec:experiments:NREL_benchmark}

The original NREL dataset \cite{nrelperformdata} includes intra-day and day-ahead forecasts.
To the best of the authors' knowledge, these are produced by NREL using a combination of weather forecasts, physical models \cite{Freeman2018_SAM}, and unspecified machine learning models.\footnote{The forecast methodology used by NREL in \cite{nrelperformdata} has not been published.}
NREL's intra-day forecasts are produced every 6 hours, with a 6-hour lead time, 6-hour horizon, and hourly granularity.
They are therefore not comparable to the short-term forecasts considered in the paper.

NREL's day-ahead forecasts are produced once a day (at midnight UTC), with an 11-hour lead time, 48-hour horizon, and hourly granularity.
While not directly comparable to the day-ahead setting considered here, which has no lead time and is updated hourly, the accuracy of NREL's day-ahead forecasts nonetheless provides a sense of the quality of the BPR forecasts.
Namely, NREL's day-ahead forecasts achieve, at the fleet level, a NMAE of 10.5\%, an RMSE of 2954MW, an ED of 37589, and a VS of 854604 and, at the asset level, NMAE of 14.4\%, an RMSE of 24.5MW, and an ED of 5469.
Overall, the data-driven BPR forecasts compare favorably in terms of NMAE, RMSE and ED at the fleet level, while NREL's asset-level forecasts achieve a slightly lower NMAE and similar RMSE and ED. Recall that NREL's forecasts use advanced physical models \cite{Freeman2018_SAM}, which are known to improve long-term forecasts. These results confirm the overall potential of BPR. 


\section{Conclusion}
\label{sec:conclusion}

The paper has proposed a novel \emph{Bundle-Predict-Reconcile} framework for hierarchical wind power forecasting, that integrates asset bundling, machine learning and forecast reconciliation techniques.
The paper has conducted extensive numerical experiments on a large industry-size dataset, thereby evaluating the impact of BPR across different forecasting tasks (short-term and day-ahead), bundling criteria, and machine learning architectures.
The results demonstrate that, compared to a baseline hierarchical setting, BPR consistently improves forecast accuracy at the fleet and asset levels, for both short-term and day-ahead forecasts.
Future work will apply BPR to probabilistic hierarchical wind power forecasting.

\bibliographystyle{IEEEtran}
\bibliography{IEEEabrv,ref}

\end{document}